\documentclass[twocolumn]{aastex631}   
\usepackage{natbib}
\usepackage{xspace}
\newcommand{\nthighlatf}{755}
\newcommand{\nthighlatbl}{1379}
\newcommand{\nthighlatu}{1208}
\newcommand{\nthighlatag}{65}
\newcommand{\ntsrc}{3407}
\newcommand{\ntrdg}{42}
\newcommand{\nthighlataa}{672}
\newcommand{\nthighlatab}{20}
\newcommand{\nthighlatac}{4}
\newcommand{\nthighlatad}{59}
\newcommand{\nthighlatba}{353}
\newcommand{\nthighlatbb}{347}
\newcommand{\nthighlatbc}{425}
\newcommand{\nthighlatbd}{254}
\newcommand{\nthighlatca}{508}
\newcommand{\nthighlatcb}{135}
\newcommand{\nthighlatcc}{117}
\newcommand{\nthighlatcd}{448}

\newcommand{\nthighlatfc}{640}
\newcommand{\nthighlatblc}{1261}
\newcommand{\nthighlatuc}{945}
\newcommand{\nthighlatagc}{50}
\newcommand{\ntsrcc}{2896}
\newcommand{\ntrdgc}{32}
\newcommand{\nthighlatcaa}{581}
\newcommand{\nthighlatcab}{18}
\newcommand{\nthighlatcac}{4}
\newcommand{\nthighlatcad}{37}
\newcommand{\nthighlatcba}{332}
\newcommand{\nthighlatcbb}{309}
\newcommand{\nthighlatcbc}{394}
\newcommand{\nthighlatcbd}{226}
\newcommand{\nthighlatcca}{397}
\newcommand{\nthighlatccb}{115}
\newcommand{\nthighlatccc}{99}
\newcommand{\nthighlatccd}{334}

\newcommand{\ntlowlat}{407}
\newcommand{\ntrdgl}{4}
\newcommand{\ntlowlatf}{37}
\newcommand{\ntlowlatbl}{79}
\newcommand{\ntlowlatu}{285}
\newcommand{\ntlowlatag}{6}
\newcommand{\ntlowlataa}{35}
\newcommand{\ntlowlatab}{0}
\newcommand{\ntlowlatac}{0}
\newcommand{\ntlowlatad}{2}
\newcommand{\ntlowlatba}{20}
\newcommand{\ntlowlatbb}{8}
\newcommand{\ntlowlatbc}{29}
\newcommand{\ntlowlatbd}{22}
\newcommand{\ntlowlatca}{78}
\newcommand{\ntlowlatcb}{12}
\newcommand{\ntlowlatcc}{10}
\newcommand{\ntlowlatcd}{185}

\newcommand{\nhighlatf}{75}
\newcommand{\nhighlatbl}{116}
\newcommand{\nhighlatu}{347}
\newcommand{\nhighlatag}{4}
\newcommand{\nsrc}{542}
\newcommand{\nrdg}{4}
\newcommand{\nhighlataa}{67}
\newcommand{\nhighlatab}{5}
\newcommand{\nhighlatac}{1}
\newcommand{\nhighlatad}{2}
\newcommand{\nhighlatba}{22}
\newcommand{\nhighlatbb}{27}
\newcommand{\nhighlatbc}{46}
\newcommand{\nhighlatbd}{21}
\newcommand{\nhighlatca}{172}
\newcommand{\nhighlatcb}{35}
\newcommand{\nhighlatcc}{41}
\newcommand{\nhighlatcd}{99}

\newcommand{\nhighlatfc}{56}
\newcommand{\nhighlatblc}{97}
\newcommand{\nhighlatuc}{273}
\newcommand{\nhighlatagc}{4}
\newcommand{\nsrcc}{430}
\newcommand{\nrdgc}{4}
\newcommand{\nhighlatcaa}{49}
\newcommand{\nhighlatcab}{5}
\newcommand{\nhighlatcac}{1}
\newcommand{\nhighlatcad}{1}
\newcommand{\nhighlatcba}{20}
\newcommand{\nhighlatcbb}{19}
\newcommand{\nhighlatcbc}{40}
\newcommand{\nhighlatcbd}{18}
\newcommand{\nhighlatcca}{130}
\newcommand{\nhighlatccb}{29}
\newcommand{\nhighlatccc}{38}
\newcommand{\nhighlatccd}{76}

\newcommand{\nlowlat}{49}
\newcommand{\nrdgl}{0}
\newcommand{\nlowlatf}{0}
\newcommand{\nlowlatbl}{1}
\newcommand{\nlowlatu}{48}
\newcommand{\nlowlatag}{0}
\newcommand{\nlowlataa}{0}
\newcommand{\nlowlatab}{0}
\newcommand{\nlowlatac}{0}
\newcommand{\nlowlatad}{0}
\newcommand{\nlowlatba}{1}
\newcommand{\nlowlatbb}{0}
\newcommand{\nlowlatbc}{0}
\newcommand{\nlowlatbd}{0}
\newcommand{\nlowlatca}{26}
\newcommand{\nlowlatcb}{4}
\newcommand{\nlowlatcc}{1}
\newcommand{\nlowlatcd}{17}

\begin{document}
\title{The Fourth Catalog of Active Galactic Nuclei Detected by the  {\em Fermi} Large Area Telescope - Data Release 3}
\author{M.~Ajello}
\affiliation{Department of Physics and Astronomy, Clemson University, Kinard Lab of Physics, Clemson, SC 29634-0978, USA}
\author{L.~Baldini}
\affiliation{Universit\`a di Pisa and Istituto Nazionale di Fisica Nucleare, Sezione di Pisa I-56127 Pisa, Italy}
\author{J.~Ballet}
\affiliation{AIM, CEA, CNRS, Universit\'e Paris-Saclay, Universit\'e de Paris, F-91191 Gif-sur-Yvette, France}
\author{D.~Bastieri}
\affiliation{Istituto Nazionale di Fisica Nucleare, Sezione di Padova, I-35131 Padova, Italy}
\affiliation{Dipartimento di Fisica e Astronomia ``G. Galilei'', Universit\`a di Padova, Via F. Marzolo, 8, I-35131 Padova, Italy}
\affiliation{Center for Space Studies and Activities ``G. Colombo", University of Padova, Via Venezia 15, I-35131 Padova, Italy}
\author{J.~Becerra~Gonzalez}
\affiliation{Instituto de Astrof\'isica de Canarias, Observatorio del Teide, C/Via Lactea, s/n, E-38205 La Laguna, Tenerife, Spain}
\author{R.~Bellazzini}
\affiliation{Istituto Nazionale di Fisica Nucleare, Sezione di Pisa, I-56127 Pisa, Italy}
\author{A.~Berretta}
\affiliation{Dipartimento di Fisica, Universit\`a degli Studi di Perugia, I-06123 Perugia, Italy}
\author{E.~Bissaldi}
\affiliation{Dipartimento di Fisica ``M. Merlin" dell'Universit\`a e del Politecnico di Bari, via Amendola 173, I-70126 Bari, Italy}
\affiliation{Istituto Nazionale di Fisica Nucleare, Sezione di Bari, I-70126 Bari, Italy}
\author{R.~Bonino}
\affiliation{Istituto Nazionale di Fisica Nucleare, Sezione di Torino, I-10125 Torino, Italy}
\affiliation{Dipartimento di Fisica, Universit\`a degli Studi di Torino, I-10125 Torino, Italy}
\author{A.~Brill}
\affiliation{NASA Postdoctoral Program Fellow, USA}
\affiliation{NASA Goddard Space Flight Center, Greenbelt, MD 20771, USA}
% ...

\begin{abstract}
An incremental version of the fourth catalog of active galactic nuclei (AGNs) detected  by the  {\em Fermi}-Large Area Telescope  is presented. This version (4LAC-DR3) derives from the third data release of the 4FGL catalog based on 12 years of  E$>$50 MeV gamma-ray data, where the spectral parameters, spectral energy
distributions (SEDs), yearly light curves, and associations have been updated for all sources. The new reported  AGNs include  587 blazar candidates and four radio galaxies. We describe the properties of the new sample and  outline changes affecting the previously published one.  We also introduce two new parameters in this release, namely the peak energy of the SED high-energy component  and the corresponding flux. These parameters allow an assessment of the Compton dominance, the ratio of the Inverse-Compton to the synchrotron peak luminosities, without relying on X-ray data.\\ 

\end{abstract}


\section{Introduction}

Since its launch in 2008, the {\em Fermi}-Large Area Telescope \citep[LAT,][]{LAT09_instrument} has enabled the discovery of new classes of gamma-ray emitters and the detection of much larger and better characterized source populations than previously achieved. Active galactic nuclei represent by far the most abundant source population  of the LAT  detected sources. The 4LAC catalog \citep[4LAC-DR1,  ][]{4LAC}, based on the 4FGL source catalog \citep[4FGL-DR1, ][]{4FGL}  established with 8 years of data,  comprised  2863   $|b|>10\arcdeg$ AGNs while 344 others were found at lower latitudes. As more data accumulate, the catalogs are regularly updated. Updates of 4FGL  use the same data version  ("P8R3") and Galactic diffuse emission model as the initial catalog. Sources previously reported are kept in  even if they fall below the TS=25 threshold over the extended period of data taking. These sources retain their original positions, in contrast to new catalogs where all positions are reevaluated and subthreshold sources are omitted. The second data releases, 4FGL-DR2 \citep{4FGLDR2} and 4LAC-DR2 \citep[comprising 285 new AGNs, ][]{4LACDR2}, were based on 10 years of data. 

Here we present the third update to the 4LAC catalog, derived from  4FGL-DR3 \citep{4FGLDR3}  using  12 years of data and comprising 1607 new sources relative to the initial 4FGL catalog. The properties of the  283 new 4FGL-DR2\footnote{Two AGNs first reported in DR2 are missing in DR3 because  the gamma-ray sources (4FGL J1242.4$-$2948 and 4FGL J1752.2$-$3002) have been either relocalised or deleted (being exceptions to the rule stated above).}  and  308 new 4FGL-DR3 AGNs (DR2 and DR3 tallies will be aggregated in the following) are discussed.  These AGNs  are all blazars except for four radio galaxies.
Besides providing a larger AGN sample for population studies, releasing periodic updates to 4LAC  offers new targets for programs  dedicated to  classifying  LAT blazars or measuring redshifts  as detections come along  \citep[e.g.,][]{Pen20,Pen21a,Pen21b,Pen21c,Des19,Raj21,CTA21}. 

The paper is organized as follows. Section 2 summarizes the analysis improvements introduced in 4FGL-DR3. Changes affecting 4LAC-DR1 AGNs are  listed in Section 3. Section 4 presents the new DR3 sources, while Section 5 discusses the peak energy of the SED high-energy component, estimated from the spectral curvature, and the derived Compton dominance. A summary closes the paper in Section 6.

\section{Analysis improvements - Source associations}

We refer the reader to the 4FGL-DR3 paper \citep{4FGLDR3} for details on the gamma-ray data analysis.
The methodology that was followed is essentially the same as that pursued in  the 4FGL-DR1 catalog. The  first stage includes the detection and localization of the sources. The second one comprises thresholding, spectral characterization and production of light curves. The same flags as in 4FGL-DR1 are then generated. They indicate  the limited robustness of the results against different analysis ingredients or warning about particular source conditions (e.g., proximity to a bright source).    The association with counterparts known at other wavelengths constitutes the final stage of the analysis. 

The (non-exhaustive) list of changes relative to 4FGL-DR1\footnote{DR2 was produced with only minor analysis changes relative to DR1.} is as follows. An updated version of the instrument-response functions (P8R3\_SOURCE\_V3) has been used.  The analysis weights have been recalculated. These weights  downplay the contribution of low-energy, low-latitude photons to the likelihood in order to reduce the associated systematic uncertainties. The handling of the energy dispersion has been improved. Bayesian priors have been applied when fitting the parameters of the diffuse emission model in each region of interest to hinder their excursion relative to their expected values (normalization =1, photon-index  correction=0).  The threshold for considering spectral curvature as significant has been lowered from 3$\sigma$ to 2$\sigma$, leading to an increase of the fraction of curved sources from 30\% in 4FGL-DR1 to 54\% in 4FGL-DR3. This change has removed unphysical upturns in the global source spectrum at  low ($<200$ MeV) and high ($>20$ GeV) energy.  It has important implications for the characterization of the LAT blazars since the peak energy of the SED high-energy component (referred to as the Inverse-Compton component in the following, assuming that relativistic electrons are the main emitting particles) and its flux can be derived from the fitted {\texttt logparabola}. The number of bins in the SED  has been increased from 7 to 8. The yearly light curves  have been updated (while the two-monthly ones reported in 4FGL-DR1 have not). Specific to the 4LAC update, the highest photon energy  detected for each source has been updated.  The information about the highest detected energy  is particularly relevant for studies of the extragalactic background light \citep[EBL, e.g., ][]{Sal21}.
 
The association  procedure makes use of two  different methods, the Bayesian method \citep{LAT10_1FGL} and the likelihood-ratio method \citep[LR,][]{LAT11_2LAC,LAT15_3LAC}, which are both based on spatial coincidence. The main change in seeking counterparts concerns the use of an updated version of the Radio Fundamental Catalog\footnote{rfc\_2021a available at \url{http://astrogeo.org/rfc/}}.  Only associations with a probability of being real  greater than 0.8  in either association method  are retained. For the 591 new AGNs, 253 are associated solely with the Bayesian method, 65 solely with the LR method and 273 with both methods. 

The same classification scheme as in the 4LAC-DR1 catalog has been followed. An optical class in terms of Flat Spectrum Radio Quasar (FSRQ) and  BL Lac-type object (BL~Lac), assessed according to the strength of the optical emission lines, is provided if spectroscopic data of sufficiently good quality are found in the literature. A SED-based class is derived from the value of the peak frequency($\nu_{\mathrm{pk}}^{\: \mathrm{syn}}$) of the synchrotron component fitted to archival data using  the SED data archive and SED(t)-Builder interactive web-tool  available at the Italian Space Agency (ASI) Space Science Data Center (SSDC)\footnote{ http://tools.ssdc.asi.it/SED/}. The class can be a low-synchrotron-peaked blazar (LSP, for sources with $\nu_{\mathrm{pk}}^{\: \mathrm{syn}} < 10^{14}$~Hz), an intermediate-synchrotron-peaked blazar (ISP, for $10^{14}$~Hz~$< \nu_{\mathrm{pk}}^{\: \mathrm{syn}} < 10^{15}$~Hz), or a high-synchrotron-peaked blazar (HSP, if $\nu_{\mathrm{pk}}^{\: \mathrm{syn}} > 10^{15}$~Hz). SED-based classification is missing for 139 new AGNs, mainly because of a lack of broad-band data.

\section{Changes to 4LAC-DR1}
For completeness,  we reiterate here the changes to 4LAC-DR1 AGNs as outlined in  the 4FGL-DR2 document \citep{4LACDR2}.       
About 200 counterpart names of DR1 sources have been changed. It was noted that  blazar names from very large surveys (such as 2MASS or WISE) were used for some 4FGL associations while more common names from radio catalogs were available.  Moreover some names referred to sources that are  offset by up to a few arcminutes from the real counterpart. We have replaced the non-radio names with those of radio counterparts whenever possible. Note that the positions reported in the 4FGL-DR1 (\texttt{RA\_Counterpart}, \texttt{DEC\_Counterpart}) fields were correct.     

Changes in associations of 4LAC-DR1 sources are listed below.
\begin{itemize}
\item Recent follow-up observations of 4FGL blazars \citep[e.g.,][]{Fup_Pen17,Fup_Pen19,Pen20,Pen21a,Pen21b,Pen21c,Des19,Raj21} have enabled the classification of 240 former  blazar candidates of unknown types (BCUs), two AGNs and two  UNKs\footnote{The UNK class corresponds to $|b|<10\arcdeg$ sources solely associated with the LR-method, which may suffer from contamination with Galactic sources.} into 214 BL Lacs and 30 FSRQs.  In particular, \cite{Pal20} found that the former BCU associated with 4FGL J1219.0+3653 is a BL~Lac with a redshift of 3.53, making it the  most  distant BL~Lac detected by the LAT.
\item The latest version of the Radio Fundamental Catalog has enabled the association with  blazar candidates of six previously unassociated sources and two SPPs (SPPs designate potential associations with supernova remnants or pulsar-wind nebulae).   These sources are 4FGL J0129.0+6312 (2MASS J01283059+6306298), 4FGL J0550.9+2552 (NVSS J055119+254909), 4FGL J0803.5+2046 (GB6 B0800+2046), 4FGL J1102.0$-$6054 (2MASS J11015838$-$6056516), 4FGL J1347.4+7309 (NVSS J134734+731812), 4FGL J1606.6+1324 (NVSS J160654+131934), 4FGL J1738.0+0236 (PKS 1735+026), and  4FGL J2249.9+0452 (WISEA J225007.35+045617.3). 
\item Three sources  (TXS 0159+085, PKS 0736$-$770, and TXS 1530$-$131) were incorrectly classified as FSRQs. They  have been reclassified as BCUs. Two other FSRQs (RX J0134.4+2638 and 2MASS J02212698+2514338) have been reclassified as BL~Lacs\footnote{TXS 0159+085 and RX J0134.4+2638 were still classified as FSRQs in 4FGL-DR3.}. 
\item Following \cite{Jar_20}, we have reclassified TXS 2116$-$077 (4FGL J2118.8$-$0723) as a Seyfert galaxy instead of a NLSY1 \citep[see ][ for an alternative view]{Pal20_2}.
\item The tentative association of 4FGL J0647.7$-$4418 with the high-mass X-ray binary  RX J0648.0$-$4418 reported in  4FGL-DR1 has been replaced by the association with the BCU SUMSS J064744$-$441946 following the multiwavelength investigation of  \citet{Assoc20_Marti}.
\end{itemize}

A total of 214 additional 4LAC-DR1 sources are now classified as  variable thanks to the extended yearly light curves produced in DR3. These sources  comprise 26 FSRQs, 105 BL~Lacs, 80 BCUs, and three radio galaxies (IC 1531, PKS 0625$-$35, and Cen A). Concerning the iconic radio galaxy Cen A, its flux dropped significantly (14\%) in the last four years spanned by this release.  

 Photons with  higher energies than  previously detected in the earlier 8 years have been found for 471 4FGL-DR1 sources  during the additional 4 years of data taking. Only E$>$10 GeV, ULTRACLEAN\_VETO photons with a probability greater than 0.95 to belong to the source have been considered.   
\begin{figure}
\centering
\includegraphics[width=0.5\textwidth]{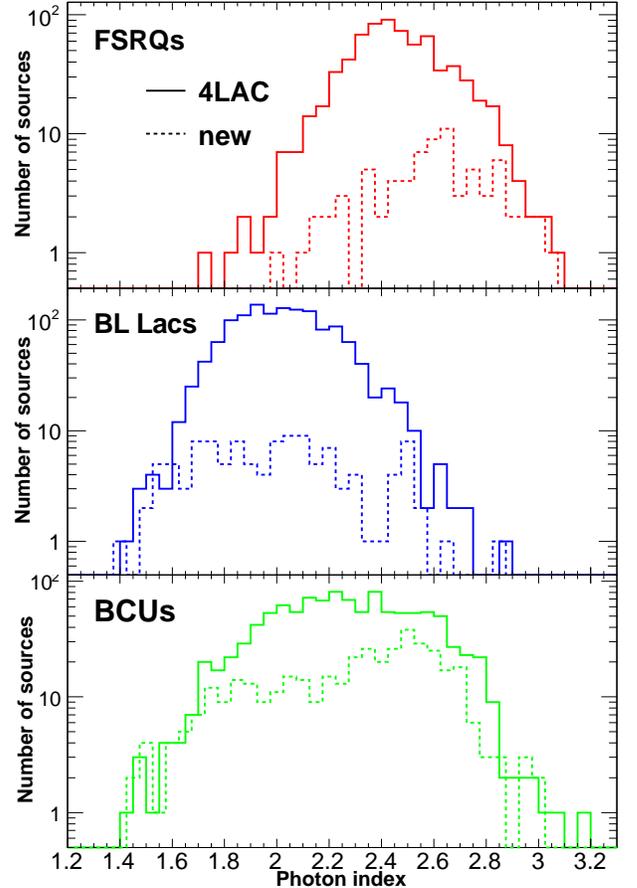}
\caption{Comparison between the photon index distributions of 4LAC-DR1 and new (DR2 and DR3) blazars for different classes. }
\label{fig:index}
\end{figure}

\begin{figure}
\centering
\includegraphics[width=0.5\textwidth]{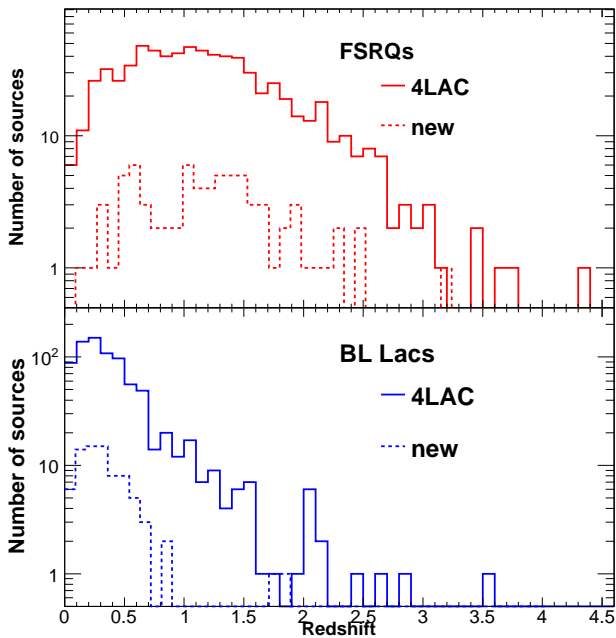}
\caption{Comparison between the redshift distributions of 4LAC-DR1 and new (DR2 and DR3) blazars for different classes. }
\label{fig:redshift}
\end{figure}
\section{The 4LAC-DR3 and low-latitude samples}

The 4LAC-DR3  sample comprises AGNs located at $|$b$|>10^\circ$, in keeping with the 4LAC defining criterion. The same defining criterion for the clean sample (i.e., sources with no analysis flags) as in 4LAC-DR1 has been used in this paper.  The AGNs lying at $|$b$|<10^\circ$ constitute the low-latitude sample.\footnote{The 4LAC-DR3 and low-latitude files are available at 
https://fermi.gsfc.nasa.gov/ssc/data/access/lat/4LACDR3/table-4LAC-DR3-h.fits 
and
https://fermi.gsfc.nasa.gov/ssc/data/access/\\lat/4LACDR3/table-4LAC-DR3-l.fits
respectively.}
Table \ref{tab:description} describes the format of the catalog FITS files.
 
The new  AGNs relative to 4LAC-DR1 include 587 blazars: 75 FSRQs,  117 BL~Lacs, 395 BCUs, and four radio galaxies (8 BCUs have been classified into 7 BL Lacs and one FSRQ since the DR2 release).  The 4LAC-DR3 comprises  542  ($|$b$|>10^\circ$) new AGNs. While FSRQs are almost evenly distributed between the northern and southern Galactic hemispheres (40 and  35 respectively), a strong deficit is observed in the South for BL~Lacs (80 vs. 37), while the opposite trend is seen for BCUs (166 vs. 229). These imbalances are probably due to fewer programs observing the southern region.  The low-latitude sample includes 49 sources (all BCUs except for one BL~Lac).

\begin{deluxetable*}{lrrr}
\tablecolumns{4}
\tabletypesize{\footnotesize}
\tablecaption{\label{tab:census_dr3}Census of new DR2-DR3 sources}
\tablewidth{0pt}
\tablehead{
\colhead{AGN type}&
\colhead{{\bf High-latitude sample}}&
\colhead{Clean Sample\tablenotemark{a}}&
\colhead{Low-latitude sample}
}
\startdata
{\bf All}&{\bf \nsrc}&{\bf \nsrcc }&{\bf \nlowlat}\\
\\
{\bf FSRQ}&{\bf \nhighlatf}&{\bf \nhighlatfc}&{\bf \nlowlatf}\\
{\ldots}LSP& \nhighlataa & \nhighlatcaa & \nlowlataa\\
{\ldots}ISP& \nhighlatab &\nhighlatcab & \nlowlatab\\
{\ldots}HSP& \nhighlatac &\nhighlatcac &\nlowlatac\\
{\ldots}no SED classification & \nhighlatad &\nhighlatcad & \nlowlatad\\
\\
{\bf BL~Lac}&{\bf \nhighlatbl}&{\bf \nhighlatblc}&{\bf \nlowlatbl}\\
{\ldots}LSP& \nhighlatba & \nhighlatcba &\nlowlatba\\
{\ldots}ISP& \nhighlatbb &\nhighlatcbb &\nlowlatbb\\
{\ldots}HSP& \nhighlatbc &\nhighlatcbc & \nlowlatbc\\
{\ldots}no SED classification & \nhighlatbd & \nhighlatcbd & \nlowlatbd\\
\\
{\bf Blazar of Unknown Type}&{\bf \nhighlatu}&{\bf \nhighlatuc}&{\bf \nlowlatu}\\
{\ldots}LSP&  \nhighlatca& \nhighlatcca & \nlowlatca\\
{\ldots}ISP& \nhighlatcb &\nhighlatccb & \nlowlatcb\\
{\ldots}HSP& \nhighlatcc &\nhighlatccc & \nlowlatcc\\
{\ldots}no SED classification & \nhighlatcd &\nhighlatccd & \nlowlatcd\\
\\
{\bf Non-blazar AGN}&{\bf \nhighlatag}&{\bf \nhighlatagc}&{\bf \nlowlatag}\\
{\ldots}RG & \nrdg & \nrdgc & \nrdgl \\
\\
\enddata
\tablenotetext{a}{Sources in the high-latitude sample without analysis flags.}
\end{deluxetable*}

\begin{deluxetable*}{lrrr}
\tablecolumns{4}
\tabletypesize{\footnotesize}
\tablecaption{\label{tab:census_tot}Census of 4LAC-DR3 sources}
\tablewidth{0pt}
\tablehead{
\colhead{AGN type}&
\colhead{{\bf High-latitude sample}}&
\colhead{Clean Sample\tablenotemark{a}}&
\colhead{Low-latitude sample}
}
\startdata
{\bf All}&{\bf \ntsrc}&{\bf \ntsrcc }&{\bf \ntlowlat}\\
\\
{\bf FSRQ}&{\bf \nthighlatf}&{\bf \nthighlatfc}&{\bf \ntlowlatf}\\
{\ldots}LSP& \nthighlataa & \nthighlatcaa & \ntlowlataa\\
{\ldots}ISP& \nthighlatab &\nthighlatcab & \ntlowlatab\\
{\ldots}HSP& \nthighlatac &\nthighlatcac &\ntlowlatac\\
{\ldots}no SED classification & \nthighlatad &\nthighlatcad & \ntlowlatad\\
\\
{\bf BL~Lac}&{\bf \nthighlatbl}&{\bf \nthighlatblc}&{\bf \ntlowlatbl}\\
{\ldots}LSP& \nthighlatba & \nthighlatcba &\ntlowlatba\\
{\ldots}ISP& \nthighlatbb &\nthighlatcbb &\ntlowlatbb\\
{\ldots}HSP& \nthighlatbc &\nthighlatcbc & \ntlowlatbc\\
{\ldots}no SED classification & \nthighlatbd & \nthighlatcbd & \ntlowlatbd\\
\\
{\bf Blazar of Unknown Type}&{\bf \nthighlatu}&{\bf \nthighlatuc}&{\bf \ntlowlatu}\\
{\ldots}LSP&  \nthighlatca& \nthighlatcca & \ntlowlatca\\
{\ldots}ISP& \nthighlatcb &\nthighlatccb & \ntlowlatcb\\
{\ldots}HSP& \nthighlatcc &\nthighlatccc & \ntlowlatcc\\
{\ldots}no SED classification & \nthighlatcd &\nthighlatccd & \ntlowlatcd\\
\\
{\bf Non-blazar AGN}&{\bf \nthighlatag}&{\bf \nthighlatagc}&{\bf \ntlowlatag}\\
{\ldots}RG & \ntrdg & \ntrdgc & \ntrdgl\\
\\
\enddata
\tablenotetext{a}{Sources in the high-latitude sample without analysis flags.}
\end{deluxetable*}

Table \ref{tab:census_dr3} gives the census of the new AGNs, while Table \ref{tab:census_tot} provides that of the whole population. Figures \ref{fig:index}  and \ref{fig:redshift} compare the photon-index and redshift distributions respectively between the new and 4LAC-DR1 samples for different blazar classes.    


The  median photon index  of the new FSRQs is  larger (2.61 vs  2.45) than that of the DR1 sample, indicating softer spectra. The median redshift is similar to 4LAC-DR1 (1.19 vs. 1.12). PKS~2318$-$087 (4FGL J2320.8$-$0823) with z=3.164 has the highest redshift of the new 4LAC-DR3 FSRQs, though four 4LAC-DR1 FSRQs have higher redshifts (up to 4.31). A total of 32 new FSRQs (7 of them with TS$>$100) are found to be variable and 12 of them, which are more significant than average,  show pronounced flares in the last 4 years of the 12-yr period. 

Of the 117 new  BL~Lacs, a SED-based  classification could be obtained for 94 of them, comprising  22 LSPs, 27 ISPs, 46 HSPs.  The new BL~Lacs  have a median photon index similar to the DR1 ones  (2.00 vs. 2.03). The median redshift of the 78 BL~Lacs with measured values is  0.28, which is  comparable to that of 4LAC-DR1 (0.34). The  maximum redshift is 0.848 for RX J1438.3+1204 (4FGL J1438.6+1205), while the maximum redshift of 4LAC-DR1 BL~Lacs is 3.53. Only 6  new BL~Lacs (all with TS$<$100) are found to be variable. 

\begin{figure}
\centering
\includegraphics[width=0.5\textwidth]{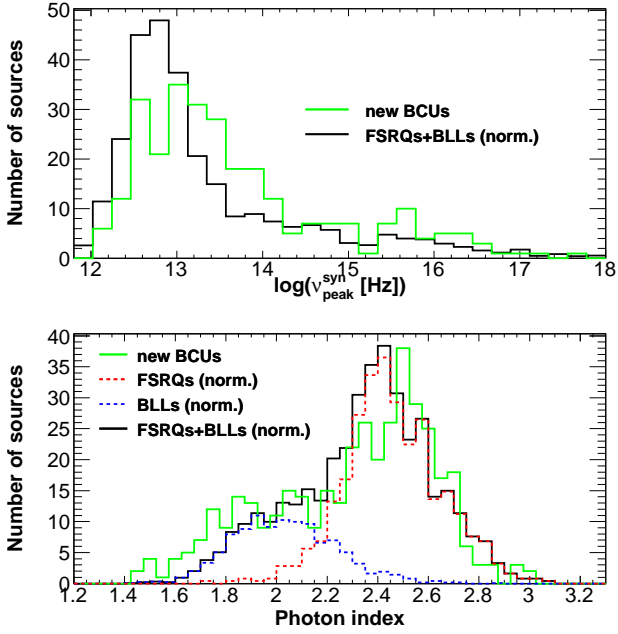}
\caption{Distributions of $\nu_{\mathrm{pk}}^{\: \mathrm{syn}}$ (top) and photon index (bottom) of the new (DR2 and DR3) BCUs (green solid) compared to the arbitrarily normalized sum of the  FSRQ and BL~Lac distributions (solid black). The relative weight between FSRQs and BL~Lacs has been multiplied by  2.5 with respect to that found in 4FGL. }
\label{fig:decomp_BCUs}
\end{figure}

\begin{figure}
\centering
\includegraphics[width=0.5\textwidth]{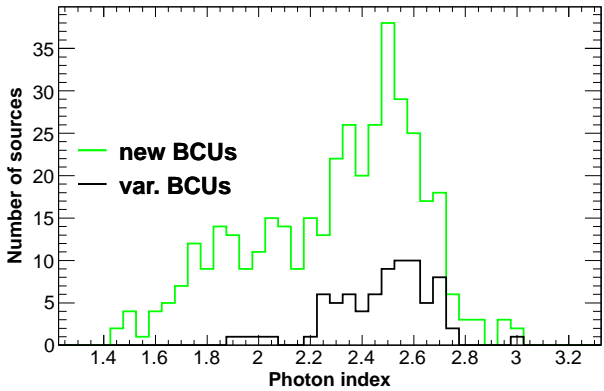}
\caption{Photon index distributions for the new BCUs and the subset of sources showing significant variability.}
\label{fig:var_BCUs}
\end{figure}

BCUs represent more than two thirds of the new blazars due to a lack of reliable spectroscopic data.  Some insight into the nature of these sources can nevertheless be gained by inspecting their photon index distributions, building on the remarkably distinct distributions exhibited by  FSRQs and BL~Lacs. The BCU distribution, with a median photon index notably higher than the corresponding DR1 value (i.e., 2.36 and 2.23 respectively), is compared to the (normalized) FSRQ and BL~Lac  distributions  in Figure \ref{fig:decomp_BCUs}. To best reproduce the BCU distribution with a linear combination of the latter distributions, the relative weight of FSRQs must be increased by a factor $\simeq$ 2.5 with respect to that found in the 4LAC-DR3 sample. In 4LAC-DR1, the BCU distribution could be well reproduced by assuming the same fractions of FSRQs and BL~Lacs as found in the classified population \citep[see Figure 6 in ][]{4LAC}.   This excess of new FSRQ candidates in 4LAC-DR3 may come from the stronger flaring activity of FSRQs relative to BL~Lacs in the LAT energy range. The photon-index distribution of the 77 variable BCUs, with 15 of them having TS$>$100, supports this idea (Figure \ref{fig:var_BCUs}). The  observation of slightly  softer spectra relative to the bulk of FSRQs  for both the new  FSRQ-like BCUs and the new FSRQs is compatible with this explanation. This  effect holds for sources of both classes with detected variability. Out of the 396 new BCUs, only 7 have measured redshifts. These have photon indices similar to those of FSRQs. This observation applies to the whole population of BCUs with measured redshifts.  

The four new radio galaxies are NGC 3078, NGC 4261, LEDA 55267, and  NGC 6454. NGC 3078 is a nearby (d=35 Mpc) compact-core-dominated galaxy \citep{Wro84}. NGC 4261 (d=30 Mpc) is a LINER Fanaroff-Riley type-I (FR I) radio galaxy, whose LAT detection was first reported by \cite{deM20}. The detection of LEDA 55267 was first reported by \cite{Pal21}. It was classified there as a Fanaroff-Riley type-0 radio galaxy, i.e., having similar  nuclear and host properties as FR I's but significantly fainter extended radio emission \citep{Gan16}. NGC 6454 is a FRI galaxy \citep{Bri08,Van12} at a distance of 130 Mpc\footnote{This source was classified as  an "AGN" in 4FGL-DR3.}.

\begin{figure}
\centering
\includegraphics[width=0.5\textwidth]{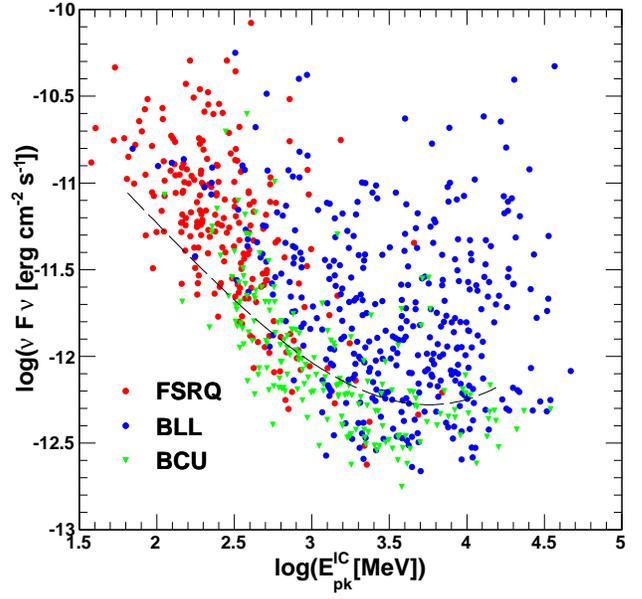}
\caption{$\nu F \nu$  of the high-energy peak  as a function of the peak energy for sources fulfilling  the conditions given in the text. Error bars have been omitted for clarity. The dashed curve corresponds to an estimated average threshold.}
\label{fig:E_nu_Fnu}
\end{figure}

\begin{figure}
\centering
\includegraphics[width=0.5\textwidth]{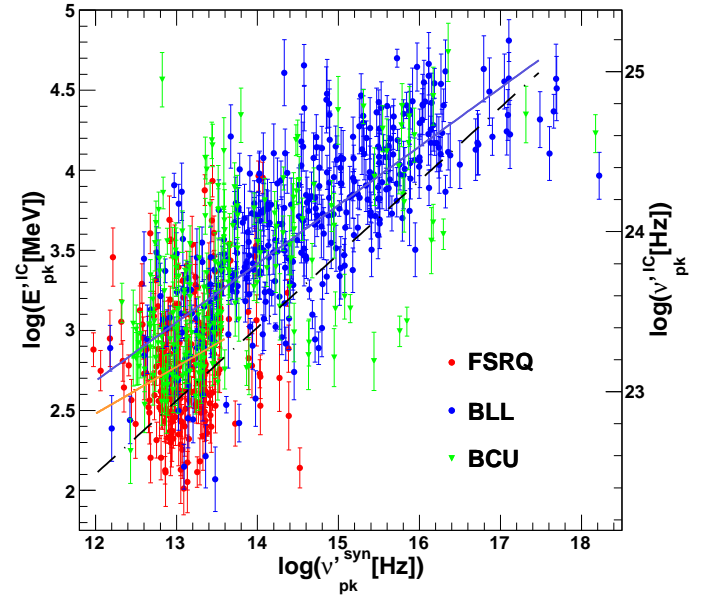}
\caption{Energy of the high-energy component plotted as a function of $\nu_{\mathrm{pk}}'^{\: \mathrm{syn}}$, both estimated  in the rest frame, for the different blazar classes. The source selection is described in the text. The blue line represents a linear fit to the BL~Lac data and the orange one (restricted to $log(\nu_{\mathrm{pk}}'^{\: \mathrm{syn}})<13.6$) to the FSRQ data. The black dash-dotted line corresponds to the prescription from \cite{SEDpaper}.}
\label{fig:nu_HE_nu_syn}
\end{figure}

\begin{figure}
\centering
\includegraphics[width=0.5\textwidth]{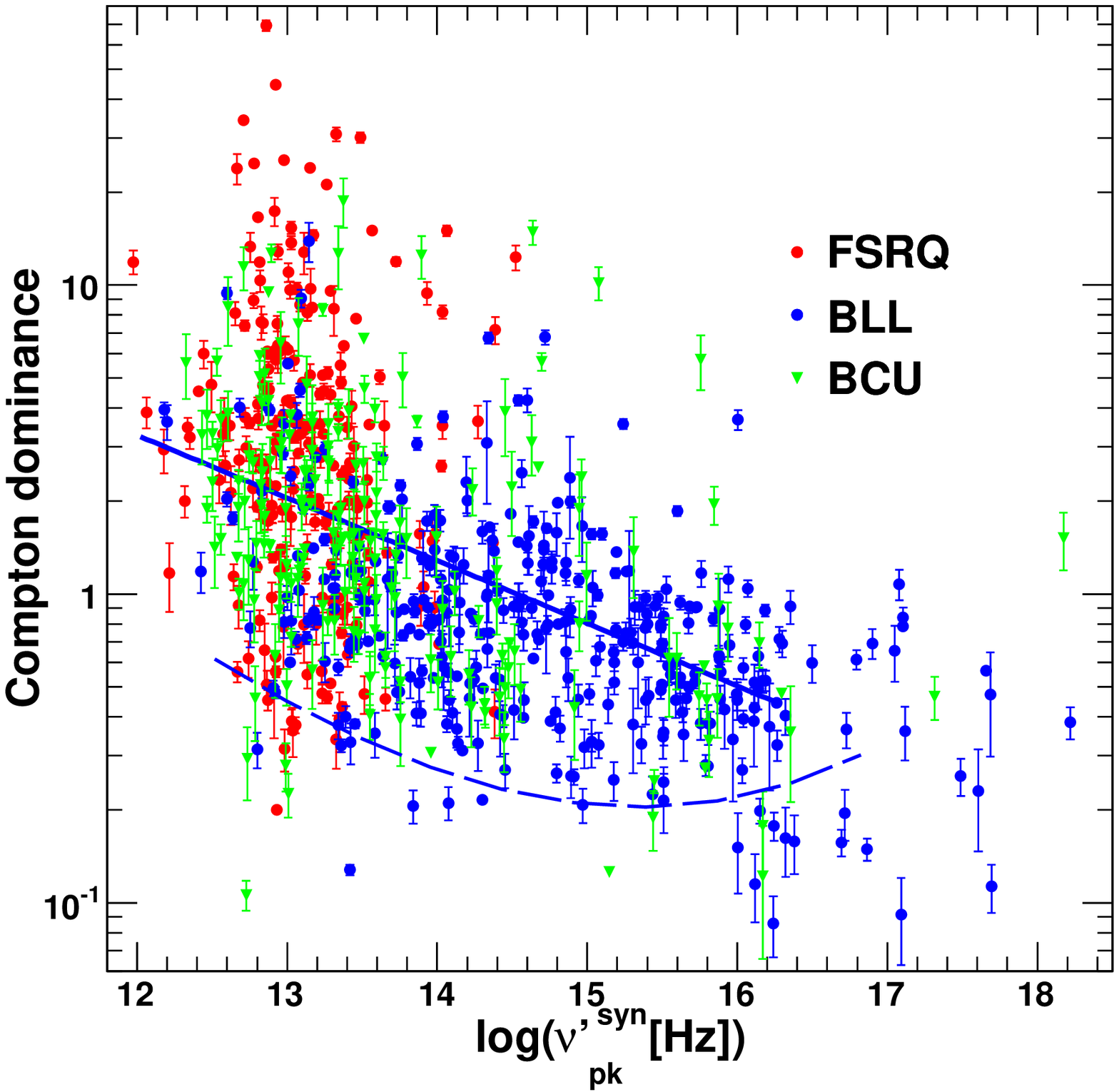}
\caption{Compton dominance plotted as a function of $\nu_{\mathrm{pk}}'^{\: \mathrm{syn}}$ for the same sources as in Figure \ref{fig:nu_HE_nu_syn}. The dashed line corresponds to an estimate of  the average threshold for BL Lacs (derived from the threshold plotted in Figure \ref{fig:E_nu_Fnu})  and the solid line to an exponential fit to the BL Lac data.}
\label{fig:CD_nu_syn}
\end{figure}
   
\section{Peak energy of the high-energy component - Compton dominance}

Two new parameters are provided in the DR3 release for the whole set of LAT blazars: the peak energy ($E_{\mathrm{pk}}^{\mathrm{IC}}$), or equivalently the peak frequency  ($\nu_{\mathrm{pk}}^{\mathrm{IC}}$), and the corresponding flux ($\nu F_{\nu}^{\mathrm{IC}}$) of the SED high-energy component  in the observer frame. Thanks to the improved statistics and the relaxed threshold for considering spectral curvature as significant,  1601 LAT blazars have a significantly curved spectrum, from which these two parameters can be estimated using a fit of a log normal function.  This assessment of $E_{\mathrm{pk}}^{\mathrm{IC}}$ is an alternative to that based on a polynomial fit to the IC  component using both X-ray and  gamma-ray data \citep[e.g., as in ][]{SEDpaper}. The latter assessment may suffer from the fact that multiple processes can contribute to the IC component, making its actual shape uncertain. 
Here we restrict the discussion to sources with a relative uncertainty on $E_{\mathrm{pk}}^{\mathrm{IC}}$ lower than 50\%.  This condition favors sources with slightly harder gamma-ray spectra than average. The effect is more pronounced for FSRQs (mean photon index difference relative to the whole sample, $\Delta\Gamma$=0.08), and diminishes for BL Lacs as $\nu_{\mathrm{pk}}^{\mathrm{syn}}$ increases (from $\Delta\Gamma$=0.05 for LSPs to $\Delta\Gamma$=0.03 for HSPs). Overall, the fraction of sources with $E_{\mathrm{pk}}^{\mathrm{IC}}$ meeting the above condition is 30\%, 28\% and 24\% for FSRQS, BL Lacs, and BCUs respectively. FSRQs preferentially exhibit $E_{\mathrm{pk}}^{\mathrm{IC}}$ below 1 GeV, while most BL Lacs have  $E_{\mathrm{pk}}^{\mathrm{IC}}$  above this value, reflecting the dichotomy seen in the photon-index distributions (Figure \ref{fig:index}). Most values of  $E_{\mathrm{pk}}^{\mathrm{IC}}$ obtained for the 9 radio galaxies meeting the above condition are notably larger than those reported from a combined fit to  the X-ray and gamma-ray data \citep[see e.g., ][]{Yas22}. An explanation for this discrepancy can be the presence of a  prominent emission from the corona/accretion disk in the  X-ray to soft gamma-ray bands, which  drives down the fitted  $E_{\mathrm{pk}}^{\mathrm{IC}}$ value in the latter case.

Except in very rare cases (5 sources),  $E_{\mathrm{pk}}^{\mathrm{IC}}$  lies within the {\em Fermi}-LAT energy range ($E_{\rm pk}>$50 MeV) and is thus reasonably  well constrained on the lower-energy side. In the following, we  apply no corrections for the extinction due to the extragalactic background light, which  affects E$>$10 GeV photons. Reported values of $E_{\mathrm{pk}}^{\mathrm{IC}}$ greater than 10 GeV must  thus be considered as lower limits, as must the associated $\nu F_{\nu}^{\mathrm{IC}}$ values.   Figure \ref{fig:E_nu_Fnu} displays  $\nu F _{\nu}^{\mathrm{IC}}$  as a function of $E_{\mathrm{pk}}^{\mathrm{IC}}$.  An estimated threshold for determining the two parameters with the required significance/accuracy is plotted in this Figure. This threshold was  calculated as a polynomial fit to the data of blazars with $\Delta E_{\mathrm{pk}}^{\mathrm{IC}}=(50 \pm 5)\%$.  

A detailed study of blazar populations is beyond the scope of this paper so we just outline some broad features.  
The synchrotron peak frequency  could be determined for 785 of these blazars, including 230 FSRQs, 362 BL~Lacs, and 193 BCUs.    
In a simple single-zone synchrotron self-Compton (SSC) model and assuming Inverse-Compton scattering in the Thomson regime, the peak Lorentz factor of the electron distribution $\gamma_{\mathrm{SSC}^{\mathrm{pk}}}$ most contributing to the electromagnetic emission can be directly assessed in a redshift-independent way  from the two peak frequencies $\gamma_{\mathrm{pk}}^{\mathrm{SSC}}=\left( \frac{3\:\nu_{\mathrm{pk}}^{\mathrm{IC}}}{4\:\nu_{\mathrm{pk}}^{\mathrm{syn}} } \right)^{1/2}$.
 These frequencies, evaluated in the source rest frame\footnote{Primed quantities are evaluated in the source rest frame. For BL Lacs and BCUs without measured redshifts, values of z=0.38 and 0.56, corresponding to the median measured redshifts for these two classes,  have been assumed respectively.}, are plotted as  a function of one another in Figure \ref{fig:nu_HE_nu_syn}.  We emphasize that the estimates of $\nu_{\mathrm{pk}}'^{\mathrm{syn}}$ and $E_{\mathrm{pk}}'^{\mathrm{IC}}$ are obtained independently, so near-empty regions in  Figure \ref{fig:nu_HE_nu_syn} do not result from an observational or analysis bias. BCUs follow the same general trend  seen for FSRQs and BL~lacs.

The correlation between  $\nu_{\mathrm{pk}}'^{\mathrm{syn}}$ and $E_{\mathrm{pk}}'^{\mathrm{IC}}$ is strong for BL~Lacs (Pearson coefficient=0.75) but weak for FSRQs (Pearson coefficient=0.08 for $log(\nu_{\mathrm{pk}}'^{\: \mathrm{syn}})<13.6$, comprising 90\% of the sample).    A fit of a linear function 
\begin{equation}
 log(E_{\mathrm{pk}}'^{\: \mathrm{IC}})= \alpha \times log(\nu_{\mathrm{pk}}'^{\: \mathrm{syn}}) +K 
\end{equation}
(with $E_{\mathrm{pk}}'^{\: \mathrm{IC}}$ in MeV and $\nu_{\mathrm{pk}}'^{\: \mathrm{syn}}$ in Hz) to the BL~Lac data provides $\alpha$=0.366  and $K=-1.71$. A similar fit for the FSRQs\footnote{Fitting the peak positions in the observer frame  yields $\alpha$=0.374 (0.344) and $K=-1.92$ ($-1.93$) for BL~Lacs (FSRQs).} restricted to  $log(\nu_{\mathrm{pk}}'^{\: \mathrm{syn}} )<13.6$ yields $\alpha$=0.293 and $K=-1.03$.   The trend obtained from  Equation 5 in \cite{SEDpaper} (derived using broad-band data from 48 bright blazars including 23 FSRQs, 23 BL~Lacs and 2 BCUs) is also shown  in Figure \ref{fig:nu_HE_nu_syn}.    

The Compton dominance  is defined as the ratio between the peak $\nu F_\nu$ for the high- and low-frequency SED components. This redshift-independent  parameter has been intensively discussed in the context of the blazar sequence  \citep[e.g., ][]{Mey11,Fin13,Nal17,Pali21}. The  Compton dominance is plotted as a function of $\nu_{\mathrm{pk}}'^{\: \mathrm{syn}}$  in  Figure \ref{fig:CD_nu_syn}, together with the threshold derived from that plotted in Figure \ref{fig:E_nu_Fnu}. 
The trend is very similar to that seen when the Compton dominance is assessed using the prescription of \cite{SEDpaper}.  Fitting the BL~Lac (log-log) data  in Figure \ref{fig:CD_nu_syn} with a linear function gives a slope of $-0.46\pm0.12$. No clear indication of an upturn around $\nu_{\rm pk}\simeq 10^{14}$ Hz manifesting the transition between Synchrotron-Self-Compton and External-Compton as the dominant emission process \citep{Fin13}  is seen in   Figure \ref{fig:CD_nu_syn}.  However, the larger scatter in Compton dominance observed for FSRQs relative to BL Lacs is likely due to this difference in main emission processes (External-Compton vs. Synchrotron-Self-Compton), in addition to different variability levels.

\section{Summary}

The new release (4FGL-DR3) includes about 19\% more blazars than the initial 4LAC-DR1, comprising  75 FSRQs,  117 BL~Lacs, 395 BCUs, and 4 radio galaxies. The large fraction of BCUs (two thirds) calls for new spectroscopic data to enable the classification.    The BCU  photon-index distribution suggests that the fraction of  FSRQs within these BCUs  is notably larger (by a factor $\simeq$2.5) than that  found in the set of classified blazars. This feature, not apparent in 4LAC-DR1, may result from a larger flaring propensity of FSRQs relative to BL Lacs in the LAT energy range, as exemplified e.g., by the very different  fractions (42\% vs. 6\%) of variable 4FGL-DR3 sources seen in these classes. Both redshift and photon-index distributions of the new  FSRQs and BL~Lacs are similar to the previously detected ones.  Thanks to the new data and a looser threshold regarding variability, 214 additional 4FGL-DR1 sources are now considered variable. A total of 1602 LAT blazars have a significantly curved spectrum, from which the peak position of the high-energy SED component and its corresponding flux  can be estimated from the gamma-ray data alone by fitting a log normal function. These parameters combined with the position of the synchrotron peak estimated from archival data allows us to derive the Compton dominance.         

\bibliography{Bibtex_4FGL_v1.bib}


\section{Acknowledgments}
Part of this work is based on archival data, software or online
services provided by the Space Science Data Center, SSDC, of the Italian
Space Agency (Agenzia Spaziale Italiana, ASI). DG and SC acknowledge
support by ASI through contract ASI-INFN 2021-43-HH.0 for SSDC, and the
Istituto Nazionale di Fisica Nucleare (INFN).

The \textit{Fermi} LAT Collaboration acknowledges generous
ongoing support from a number of agencies and institutes that have supported
both the development and the operation of the LAT as well as scientific data
analysis.  These include the National Aeronautics and Space Administration and
the Department of Energy in the United States, the Commissariat \`a l'Energie
Atomique and the Centre National de la Recherche Scientifique / Institut
National de Physique Nucl\'eaire et de Physique des Particules in France, the
Agenzia Spaziale Italiana and the Istituto Nazionale di Fisica Nucleare in
Italy under ASI-INFN Agreements No. 2021-43-HH.0, the Ministry of Education, Culture, Sports, Science and Technology
(MEXT), High Energy Accelerator Research Organization (KEK) and Japan
Aerospace Exploration Agency (JAXA) in Japan, and the K.~A.~Wallenberg
Foundation, the Swedish Research Council and the Swedish National Space Board
in Sweden. Additional support for science analysis during the operations phase is gratefully acknowledged from the Istituto Nazionale di Astrofisica in Italy and the Centre National d'\'Etudes Spatiales in France.
This work performed in part under DOE Contract DE-AC02-76SF00515.

This research has made use of the \citet{NED1} which is operated by the Jet Propulsion Laboratory, California Institute of Technology, under contract with the National Aeronautics and Space Administration, and of archival data, software and online services provided by the ASI Space Science Data Center (SSDC) operated by the Italian Space Agency.


\appendix
\section{Description of the FITS version of the 4LAC-DR3 catalog}
\setcounter{table}{0}
\renewcommand{\thetable}{A\arabic{table}}
Table A1 provides a description of the  FITS catalogs available in the online Journal. There are separate FITS catalogs for the  (high-latitude)  4LAC-DR3 and low-latitude sources.
\begin{deluxetable*}{lccl}
\setlength{\tabcolsep}{0.04in}
\tablewidth{0pt}
\tabletypesize{\scriptsize}
\tablecaption{4LAC-DR3 FITS Format \label{tab:description}}
\tablehead{
\colhead{Column} &
\colhead{Format} &
\colhead{Unit} &
\colhead{Description}
}
\startdata
Source\_Name & 18A & \nodata & Source name 4FGL JHHMM.m+DDMMa\tablenotemark{a} \\
DataRelease  & I &  & 1 for 4FGL, 2 for new in DR2, 3 for new or changed in DR \\
RAJ2000 & E & deg & Right Ascension \\
DEJ2000 & E & deg & Declination \\
GLON & E & deg & Galactic Longitude \\
GLAT & E & deg & Galactic Latitude \\
Signif\_Avg & E & \nodata & Source significance in $\sigma$ units over the 50~MeV to 1~TeV band \\
Flux1000 & E & cm$^{-2}$ s$^{-1}$ & Integral photon flux from 1 to 100~GeV \\
Unc\_Flux1000 & E & cm$^{-2}$ s$^{-1}$ & $1\sigma$ error on integral photon flux from 1 to 100~GeV \\
Energy\_Flux100 & E & erg cm$^{-2}$ s$^{-1}$ & Energy flux from 100~MeV to 100~GeV obtained by spectral fitting \\
Unc\_Energy\_Flux100 & E & erg cm$^{-2}$ s$^{-1}$ & $1\sigma$  error on energy flux from 100~MeV to 100~GeV \\
SpectrumType & 17A & \nodata & Spectral type in the global model (PowerLaw, LogParabola, \\
& & &  PLSuperExpCutoff) \\
PL\_Index & E & \nodata & Photon index when fitting with PowerLaw \\
Unc\_PL\_Index & E & \nodata & $1\sigma$ error on PL\_Index \\
Pivot\_Energy & E & MeV & Pivot Energy \\
LP\_Index & E & \nodata & Photon index at Pivot\_Energy ($\alpha$) when fitting with LogParabola \\
Unc\_LP\_Index & E & \nodata & $1\sigma$ error on LP\_Index \\
LP\_beta & E & \nodata & Curvature parameter ($\beta$) when fitting with LogParabola \\
Unc\_LP\_beta & E & \nodata & $1\sigma$ error on LP\_beta \\
Flags & I & \nodata & Analysis flags \\
CLASS & 6A & \nodata & Class designation for associated source \\
ASSOC1 & 30A & \nodata & Name of identified or likely associated source \\
ASSOC\_PROB\_BAY & E & \nodata & Probability of association according to \\
& & & the Bayesian method \\
ASSOC\_PROB\_LR & E & \nodata & Probability of association according to \\
& & & the Likelihood Ratio method \\
Counterpart\_Catalog & 10A & & Counterpart catalog driving the association \\ 
RA\_Counterpart & D & deg & Right Ascension of the counterpart ASSOC1 \\
DEC\_Counterpart & D & deg & Declination of the counterpart ASSOC1 \\
Unc\_Counterpart & E & deg & 95\% precision of the counterpart localization \\
VLBI\_Counterpart   & 14A &  \nodata & Name of the VLBI counterpart \\ 
Redshift   &  E & \nodata  & Redshift \\
SED\_class  &  6A & \nodata & SED-based class\\
HE\_EPeak & E & MeV & Energy in the observer frame of the high-energy SED peak \\
Unc\_HE\_EPeak & E & MeV &  1$\sigma$ error on energy of the high-energy SED peak\\
HE\_nuFnuPeak & E & erg cm$^{-2}$ s$^{-1}$ & $\nu F \nu$ at high-energy-peak frequency \\
Unc\_HE\_nuFnuPeak & E & erg cm$^{-2}$ s$^{-1}$ &  1$\sigma$ error on spectral energy distribution at high-energy-peak frequency \\
nu\_syn &  E & Hz & synchrotron-peak frequency  in the observer frame \\
nuFnu\_syn  & E  & erg cm$^{-2}$ s$^{-1}$ &  $\nu F\nu$  at synchrotron-peak frequency \\
Variability\_Index & E &  \nodata & Variability index \\
Frac\_Variability  & E &  \nodata & Fractional variability  \\
Unc\_Frac\_Variability   & E &  \nodata & 1$\sigma$ error on fractional  variability \\
Highest\_energy &  E & GeV &  energy  (if greater than 10 GeV) of the highest-energy ULTRACLEANVETO photon \\
& & & with association probability $P>0.95$ \\
\enddata
\tablenotetext{a}{The coordinates are rounded, following the International Astronomical Union convention.}
\end{deluxetable*}

\end{document}